\documentclass[12pt,preprint]{aastex}   % for single-column preprint

\newcommand{\Msol}{\mbox{$M_\odot$}}
\newcommand{\Vlsr}{\mbox{$V_{\rm LSR}$}}
\newcommand{\kms}{\mbox{km s$^{-1}$}}
\newcommand{\beam}{\mbox{beam$^{-1}$}}

\newcommand{\minus}{\mbox{$-$}}
\newcommand{\xone}{\mbox{$x_{1}$}}
\newcommand{\xtwo}{\mbox{$x_{2}$}}

\newcommand{\pv}{\mbox{$p$--$v$}}

\shorttitle{Molecular Gas around Double Nucleus in M83}
\shortauthors{Sakamoto et al.}

\slugcomment{Accepted for publication in {\it The Astrophysical Journal Letters}}

\begin{document}

\title{Molecular Gas around the Double Nucleus in M83}

\author{Kazushi Sakamoto\altaffilmark{1}, 
	Satoki Matsushita\altaffilmark{2},
	Alison B. Peck\altaffilmark{1}, \\
	Martina C. Wiedner\altaffilmark{3,1}, and
	Daisuke Iono\altaffilmark{1,4} 
	}

\altaffiltext{1}{Harvard-Smithsonian Center for Astrophysics,
SMA, 645, N. A'ohoku Place, Hilo, HI 96720}
\altaffiltext{2}{Academia Sinica, Institute of Astronomy and Astrophysics, 
P.O. Box 23-104, Taipei 106, Taiwan}
\altaffiltext{3}{Physikalisches Institut, Universit\"at zu K\"oln, Z\"ulpicher Str. 77,
50937 K\"oln, Germany}
\altaffiltext{4}{Department of Astronomy, University of Massachusetts, Amherst, MA 01002}

%%%%%%%%%%%%%%%%%%%%%%%%%%%%%%%%%%%%%%%%%%%%%%%%%%%%%
\begin{abstract}
The center of M83, a barred starburst galaxy with a double nucleus, has been observed 
in  the CO($J$=2--1) and CO($J$=3--2) lines with the Submillimeter Array.
The molecular gas shows a distribution and kinematics  typical for barred galaxies at 
$\sim$kpc radii, but reveals unusual kinematics around the double nucleus in
the central $\sim$300 pc.
Our CO velocity data show that the visible nucleus in M83 is at least 3\arcsec\ (65 pc)
away from the galaxy's dynamical center, which most likely coincides with the
center of symmetry previously determined in $K$ band and is suggested to host
another nucleus.
We discovered high-velocity molecular gas associated with the
visible off-center nucleus, and also found a steep velocity gradient across it. 
We attribute these features to a gas disk rotating around the off-center nucleus,
which may be the remnant  of a small galaxy accreted by M83.
The dynamical mass of this component is estimated to be 
$3\times 10^8 \Msol$ within a radius of 40 pc.
The dynamical perturbation from the off-center nucleus may have played a key role 
in shaping the lopsided starburst.
\end{abstract}


%%%%%%%%%%%%%%%%%%%%%%%%%%%%%%%%%%%%%%%%%%%%%%%
\section{Introduction}

M83 (NGC 5236) is a nearby face-on barred spiral galaxy 
with a double nucleus and a nuclear starburst
($D=4.5$ Mpc, $1\arcsec = 22$ pc; Thim et al. 2003).
The galaxy is  
exceptionally complex in the central 300 pc,
despite the symmetric appearance of its bar and spiral arms. 
In $K$ band, the brightest point, called the `visible nucleus', 
is offset by  3\farcs4  (75 pc) from
the centroid (i.e., center of symmetry) determined from the isophotes at radii of
0.3 -- 0.7 kpc \citep[hereafter TTG]{Thatte00}. 
The centroid suffers from large extinction, and does not show a peak in $K$ band.
However, the stellar velocity dispersion peaks both at the visible nucleus and 
near the isophotal centroid, and suggests a mass of $\sim$10$^{7}$\Msol\ for each.
This suggests a double nucleus (TTG).
The visible nucleus has a hard power-law spectrum in X-ray due 
either to an accreting supermassive black hole or to X-ray binaries \citep{Soria02}. 
\citet{Elmegreen98} found a 
double circumnuclear ring in their $J-K$ color index map.
The  `inner' and `outer' rings of 190 pc and 60 pc radius are not concentric.
The galaxy hosts a circumnuclear starburst 
mainly in the `starburst arc' of $\sim$250 pc length
that lies between the two rings \citep{Gallais91, Harris01}.
The galactic center has a large amount of molecular gas,
estimated to be $10^{7.5\mbox{--}8.5}\Msol$ in $r\leq300$ pc \citep{Handa90,Israel01}.

The proximity, face-on configuration, abundant molecular gas, starburst, and double nucleus 
make
M83 one of the most interesting targets 
to study spiral galaxies and their nuclear activity through molecular gas. 
We observed M83 as a part of the early science program of the Submillimeter Array
(SMA)\footnote{
The Submillimeter Array is a joint
project between the Smithsonian Astrophysical Observatory and the
Academia Sinica Institute of Astronomy and Astrophysics, and is
funded by the Smithsonian Institution and the Academia Sinica.
}. 
The new telescope in Hawaii can easily observe the galaxy at $\delta=-30\degr$.
In this {\it Letter}, we first describe the observations and the bar-driven gas dynamics
in the central 2 kpc,  and then show that the visible nucleus is indeed offset from the dynamical center
of M83 and has a velocity feature indicative of its own gas disk. 
We discuss the origin of the double nucleus, and its relation to the starburst activity.

%%%%%%%%%%%%%%%%%%%%%%%%%%%%%%%%%%%%%%%%%%%%%%%
\section{SMA Observations}

M83 was observed with the SMA \citep{Ho04} 
from February through May 2003 during the commissioning of the array.
We observed CO($J$=2--1) and CO($J$=3--2) lines in the galactic center 
with a total on-source time of 14 hr and 10 hr, respectively.
Five of the eight SMA antennas were available, 
except for the first 230 GHz observations which had only four antennas.
A partial reconfiguration of the array during our observing period 
gave us 14 independent baselines for each band.
Zenith opacity was between 0.03 and 0.15 at 225 GHz
during our observations, and DSB system temperatures
toward the galaxy
were typically 150 K at 1.3 mm and 500 K at 0.85 mm.
The correlator bandwidth and resolution were 640 MHz and 0.8125 MHz,
respectively.
The CO(2--1) observations were the first mosaic imaging with the SMA.
Three positions
separated by 25\arcsec\ from each other along the bar were cycled through 
every 10 minutes.
We calibrated the system gain every $\sim$30 minutes by observing
a pair of nearby quasars, J1337\minus129 and J1313\minus333 at 1.3 mm
and J1337\minus129 and 3C279 at 0.85 mm.
Interferometric pointing was made during the observations toward
the gain calibrators to keep pointing errors $\lesssim 5\arcsec$.
The passband was calibrated using Jupiter or Saturn,
and the flux scale using Uranus.
The primary calibrator J1337\minus129 dimmed by $\sim$30 \% in 1.3 mm
in our observing period of 76 days.

        Calibration of the raw data was made using MIR, 
and the imaging and data analysis were made with MIRIAD and AIPS.
Channel maps were made every 10 \kms\ and then combined to make
moment maps. No continuum was detected from channels free of
line emission, with 5 $\sigma$ upper limits of 50 and 150 mJy \beam\
for 1.3 mm and 0.85 mm, respectively. 
Our interferometric maps recovered 90 \% and 30--50 \% of the total flux
at the galactic center in CO(2--1) and CO(3--2), respectively, based on a
comparison with single-dish observations  
\citep{Crosthwaite02, Dumke01, Israel01}.

%%%%%%%%%%%%%%%%%%%%%%%%%%%%%%%%%%%%%%%%%%%%%%%

\section{Molecular Gas in the Central 2 kpc}
\begin{figure}
\epsscale{0.57}
\plotone{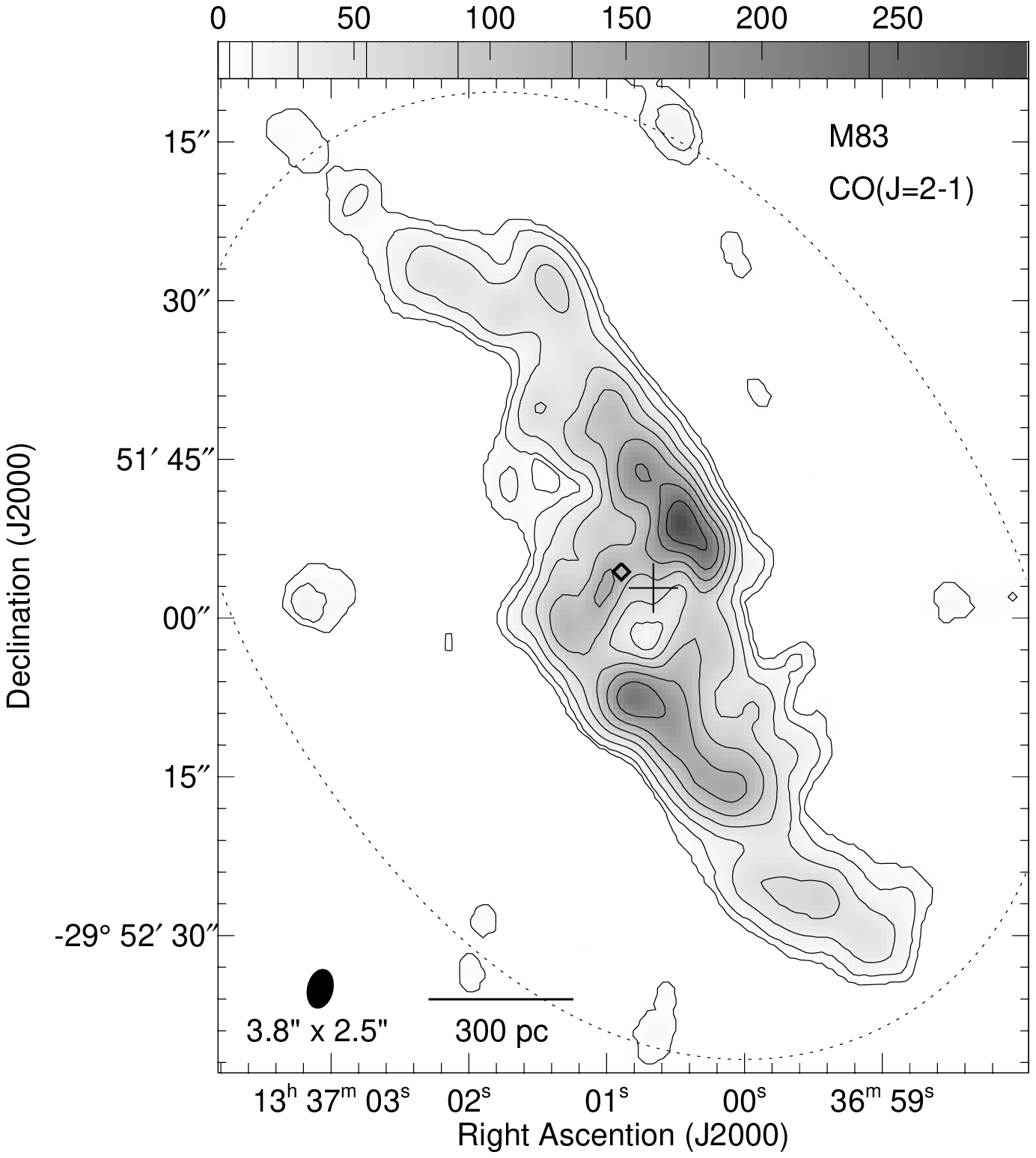} \\
\epsscale{0.57}
\plotone{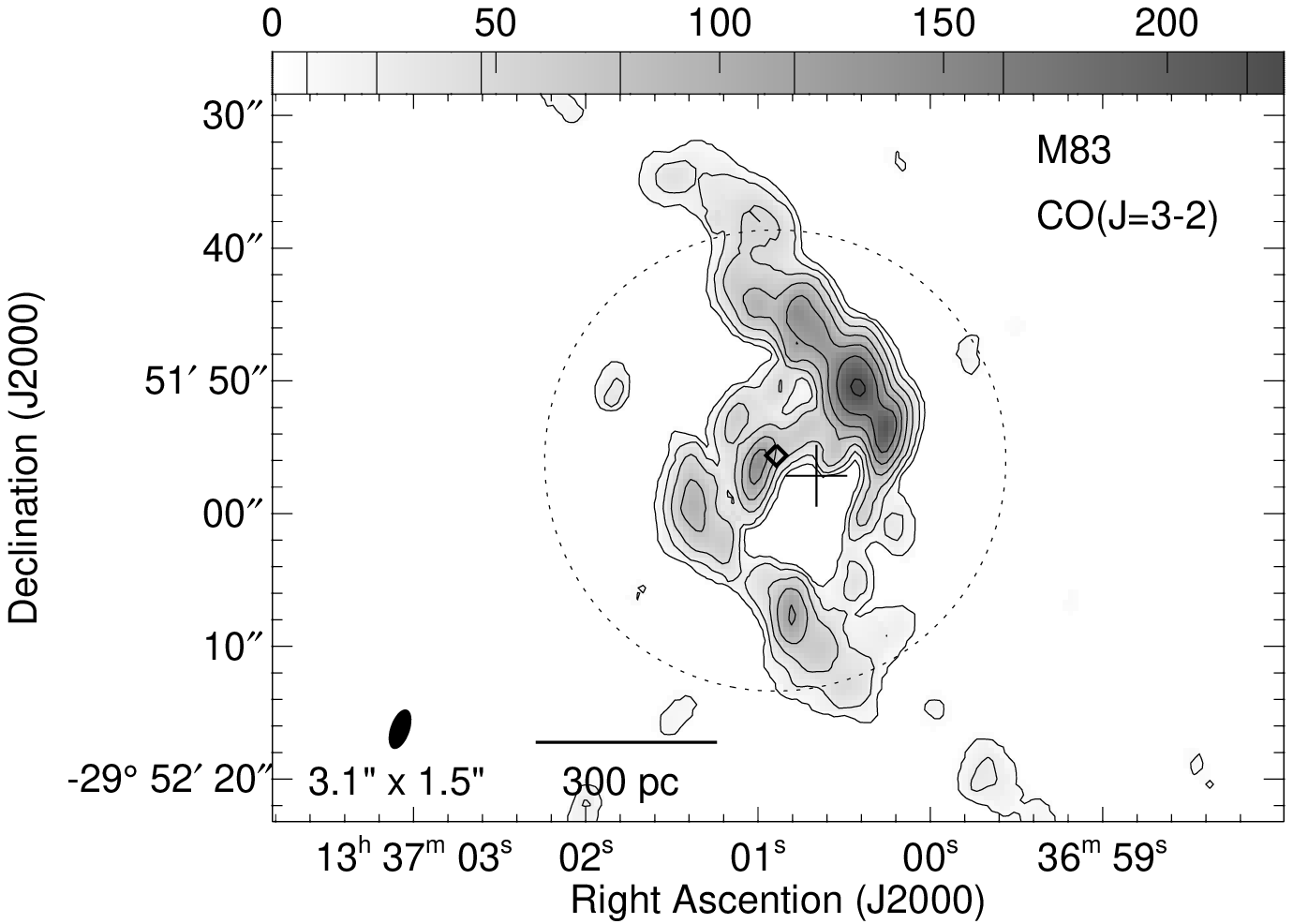}
\caption{Integrated intensity maps of CO(2--1) and CO(3--2) in the center of M83. 
The diamond marks the visible nucleus and the cross is at the isophotal centroid
in $K$ band.
The synthesized beam of each map is shown at the bottom left corner.
The unit of intensity is Jy \beam\ \kms.
The dotted contour in the CO(2--1) map shows where the power pattern of the mosaicked primary beam
drops to 30 \% of its peak. 
The one in the CO(3--2) map shows the half power width of the primary beam.
The maps are not corrected for the power (i.e., sensitivity) patterns.
\label{fig.mom0} 
}
\end{figure}

The CO(2--1) and CO(3--2) integrated intensity maps (Figure \ref{fig.mom0})
reveal two gas ridges at the leading side of the stellar bar (p.a.  $\sim$50\degr),
and a nuclear gas ring of  $\sim$300 pc diameter. 
There are additional features inside the ring, which we discuss in \S \ref{s.nucleus}.
The overall morphology of molecular gas agrees well with that in the $J-K$ ($\approx$ extinction) image 
of Elmegreen et al. (1998); for example, the $\sim$300 pc nuclear ring is their
outer ring.

\placefigure{fig.mom0}

\begin{figure}
\epsscale{0.7}
\plotone{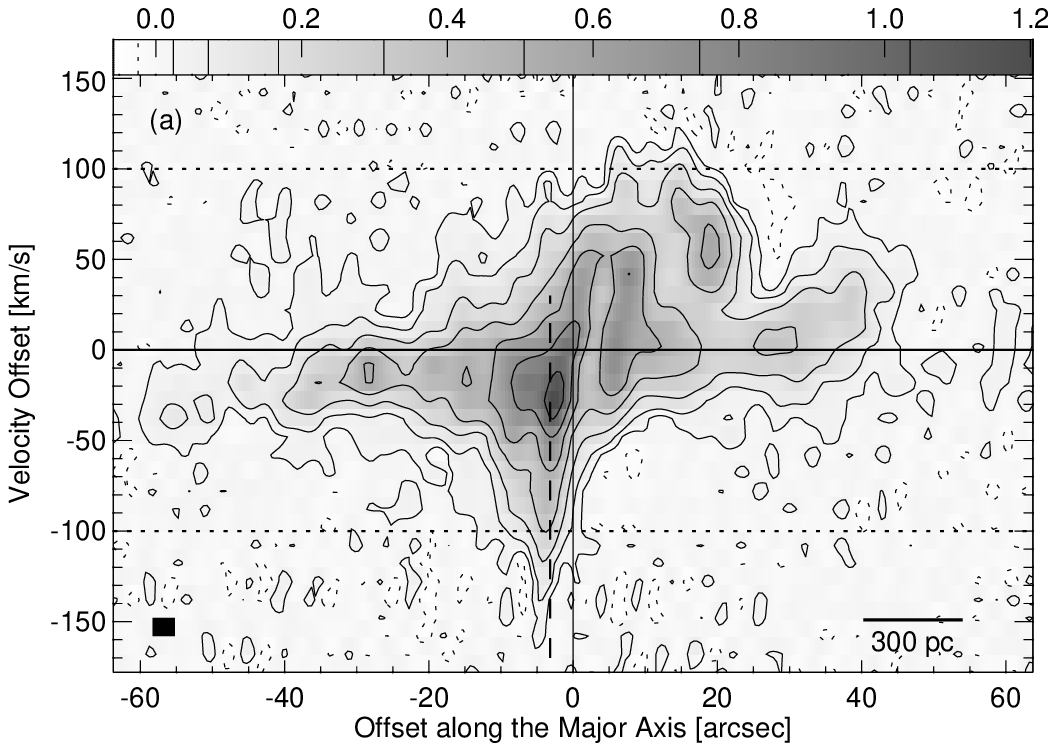} \\
\epsscale{0.7}
\plotone{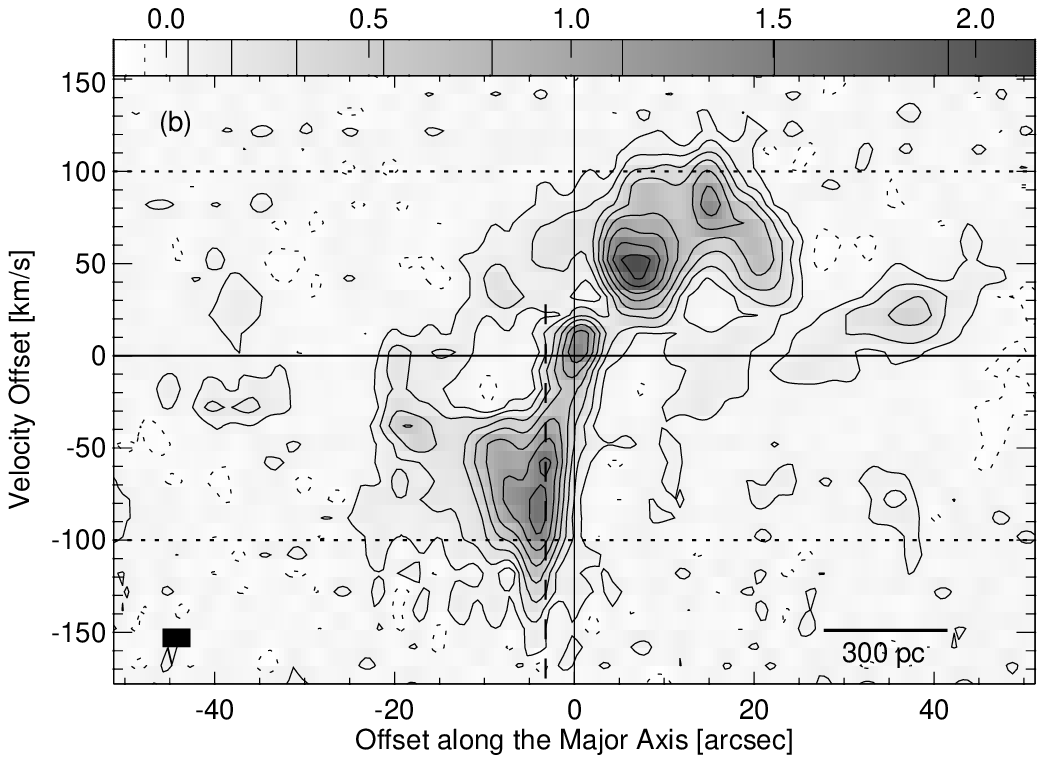}
\caption{CO(2--1) position-velocity diagrams along the major axis  (P.A.$=226\degr$). 
Position is measured from the isophotal centroid and velocity is from the
systemic velocity of 508 \kms.  
The intensity unit  is Jy \beam.
The black rectangles on the bottom-left corners show spatial and velocity resolutions.
The dashed lines at $-3\farcs2$ mark the position of the visible nucleus.
(a) The cut (or `slit') is 30\arcsec\ wide and includes most emission in our mosaic.
(b) The cut is 5\arcsec\ wide, with its central axis 1\arcsec\ offset from the major 
axis so that it  crosses the visible nucleus.
Most of the linear gas ridges lie outside this cut.
\label{fig.pv}}
\end{figure}

The pair of ridges and the nuclear ring must be due to 
the two types of oval orbits in the bar.
Namely, the former must be on the \xone\ orbits, which are the
dominant type of orbits in a bar and are elongated along the bar, 
while the latter must be on the \xtwo\ orbits, which are seen near the center of a bar and are
elongated perpendicular to the bar \citep{Athanassoula92}.
The presence of two major kinematical components is evident in the position-velocity diagram
of the CO(2--1) emission, Fig. \ref{fig.pv}(a). 
One component that runs throughout our field of view with slow rotation
is from the gas ridges, 
while the other with faster rotation near the center 
is from around the nuclear ring ($r \approx 8\arcsec$). 
Our viewing angle of M83, from which the line of nodes coincides with the bar, makes
gas on  \xone\  orbits appear slower and  \xtwo\ gas appear faster.

\placefigure{fig.pv}

Besides the ridges and the nuclear ring, there is faint emission between them 
in the CO(2--1) map Fig. \ref{fig.mom0}(a).
This weak emission and part of the gas ridges form a parallelogram.
The morphology and location of this component suggest that it is on
oval orbits whose major axes are neither perpendicular nor parallel to the bar but in-between.
Indeed, it is theoretically expected that gaseous orbits 
between  \xone\ and \xtwo-dominated regions
continuously change their
major axes from parallel to the bar to perpendicular to the bar as the orbital radius
decreases \citep{Wada94, Lindblad94}.
In addition, gas funneled in the bar through one of the gas ridges on \xone\ orbits can 
eventually collide with gas in the nuclear ring on \xtwo\ orbits,
and can be `sprayed back' onto the other ridge to form a parallelogram
\citep{Binney91, Regan99}.

%%%%%%%%%%%%%%%%%%%%%%%%%%%%%%%%%%%%%%%%%%%%%%%

\section{Molecular Gas in the Central 300 pc} \label{s.nucleus}
\begin{figure}
\epsscale{0.7}
\plotone{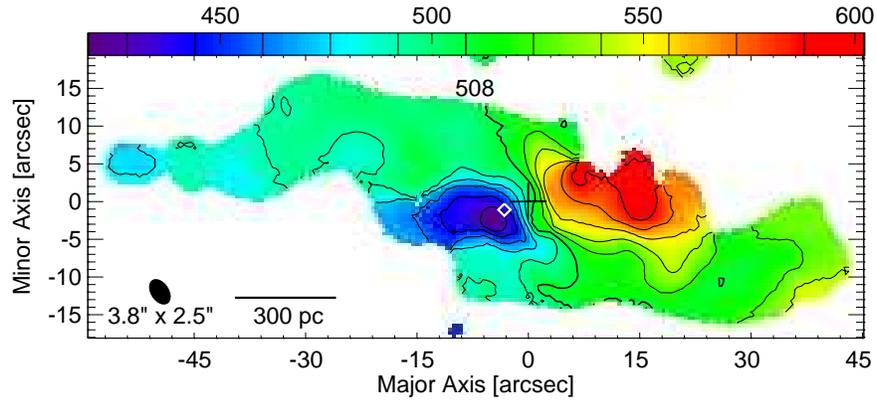}
\caption{CO(2--1) velocity map of M83. 
Contours are in 16 \kms\ steps from 428 \kms\ (\Vlsr).  
The systemic velocity of 508 \kms\ is shown by a thick contour.
The diamond at ($-3\farcs2,-1\farcs1$) is the visible nucleus and the cross at ($0,0$) 
is the $K$-band isophotal centroid.  
The map is rotated to make the major axis horizontal.
\label{fig.mom1}}
\end{figure}

We established kinematically the offset of the visible nucleus from the dynamical center 
in the following way.
The systemic velocity of the galaxy was first determined to be $508\pm3$ \kms\ (\Vlsr) from
the  CO(2--1) mean velocity data (Figure \ref{fig.mom1}) 
by searching for the velocity whose contour is least biased toward either blue- and red-shifted direction. 
The velocity map was not fitted to derive the systemic velocity because 
the emission covered only a small area and there is noncircular motion
due to the bar\footnote{
We adopt the position angle of 226\degr\ for the
receding major axis of the galaxy, based on the CO and HI velocity fitting
across the disk \citep{Crosthwaite02}. 
In Figure \ref{fig.mom1}, the contour for the systemic velocity 
has an angle of  $\sim$75\degr\ rather than 90\degr\ to this major axis.
Noncircular motion in the bar is probably most responsible for this,
though warp of the disk could be partly responsible.
}
and to the visible nucleus as discussed below. 
The dynamical center should lie on 
the systemic-velocity contour.
The $K$ band center determined by TTG from isophotes at
radii between 12\arcsec\  and 30\arcsec,  
shown as a cross in the velocity map, is almost on that contour.
Thus, it is most likely that the dynamical center of M83 is on, or in the vicinity of, the
isophotal centroid, as morphologically expected.
On the other hand, the visible nucleus shown as a diamond in the velocity map is
clearly off the systemic-velocity contour and must be
at least 3\arcsec\ (65 pc) away from the dynamical center.
\placefigure{fig.mom1}

We discovered high-velocity gas around the off-center, visible nucleus.
In Figure \ref{fig.pv}, where the off-center nucleus is at $-3\farcs2$
while the isophotal centroid is at 0\arcsec,
the \pv\ diagrams 
show a high-velocity component at around $ -5\arcsec$ and $-160$ \kms. 
There is no counterpart in the receding side; the largest
velocity at $+5\arcsec$ is $+120$ \kms.
Spatially, the component is compact and is within 2\arcsec\ of the off-center nucleus
at the most blueshifted velocity. 
This compact high-velocity component is detected in both CO(2--1) and CO(3--2). 
It shows up in the integrated intensity maps as the peak 
$\sim$2\arcsec\ southeast of the visible nucleus.
The large blueshift of this component is also evident in the
mean velocity map (Fig. \ref{fig.mom1}).
The position-velocity diagram in Fig. \ref{fig.pv}(b) clearly shows 
that the high-velocity gas is a part of the
continuous ridge of emission that has
a large velocity gradient across the visible, off-center nucleus. 
The velocity gradient around the nucleus is roughly 
along the galaxy's major axis, as seen in the mean velocity map.

We interpret this feature as a gas disk that rotates around the off-center nucleus
but not directly around the dynamical center of M83.
The feature is unlikely to be due to streaming motion in the bar potential,
though molecular gas in the center of barred galaxies often shows  a large velocity gradient
due to noncircular motion  \citep{Sakamoto00}.
This is because the feature is seen only on one side of the dynamical center and only
in blueshifted velocities, and also because the visible nucleus is at the center of the feature.
The feature is also unlikely to be due to an outflow.
It is because both starburst and a nuclear jet, 
the latter of which may exist along the minor axis of the galaxy \citep{Cowan94},
would cause a velocity gradient along the minor axis rather than the major axis as we see here.

The extent and the total width of the line-of-sight velocity of this component are
 80 pc and 140 \kms, or 120 pc and 180 \kms\ if we include the peak
near (+1\arcsec, +10\kms).  
The size and location of the gas disk are roughly consistent with those of the inner 
ring in Elmegreen et al. (1998).
The dynamical mass of the disk is $3\times 10^8 \Msol$ within a radius of 40 pc
assuming the same inclination as M83 (25\degr); 
it would be a factor of 4 smaller if the inclination was 60\degr. 
The mass distribution in the disk is extended, and not dominated by a single compact object,
because the mass within 6.5 pc of the visible nucleus  was estimated to be 
1.6$\times 10^{7}$\Msol\ (TTG).

What are the off-center nucleus and the gas rotating around it?
TTG concluded that the visible nucleus could not be a young star cluster,
but had to be a dynamically hot nucleus, 
on the basis of its high velocity dispersion
and stellar population.
This conclusion is consistent with the observation of \citet{Harris01} that the mass
of the individual young clusters in the nuclear region of M83 (mostly on the
starburst arc) is less than about $10^{5}\Msol$. 
The eccentric disk model used for the double nucleus in M31\citep{Tremaine95} 
is also dismissed for M83 because, among other things,
the rotation of stars and gas around the dynamical center is not Keplerian.

A plausible model for the visible off-center nucleus is 
that it is the remnant core of a small galaxy that merged into M83. 
The prototype major merger Arp 220 has two surviving nuclei at a 
projected separation of $\sim$300 pc with gas disks
rotating around each of them \citep{Sakamoto99, Genzel01}. 
A less violent minor merger may have produced the configuration seen here.
M83 has a faint stellar arc of $\sim$30 kpc length indicating 
the accretion of at least one satellite \citep{Malin97}.
If the visible nucleus came from outside, then it does not need to be on the disk
plane of M83. 
However,  the nucleus is probably not far away from the mid-plane of M83 because
(1) the chances of seeing it at the small projected distance from
the dynamical center are low if the nucleus is far from  the disk plane,
(2) the mean velocity of gas and stars of the visible nucleus is roughly on the rotation curve of M83,
and 
(3) numerical simulations of satellite accretion show that a satellite 
rapidly looses its orbital radius  after it settles on the disk of the larger galaxy,
since disk stars are the main source of torque on the satellite
\citep{Walker96}.
If the accreted nucleus has spent a long time in the
galactic plane of M83, then the gas rotating around the nucleus, presumably as a mini-disk, 
may not be from the original satellite 
but may have been  captured from M83 during the accretion.
Regarding the large extinction toward the other nucleus at the dynamical center
(presumably the true nucleus of M83),
it is conceivable that 
the mini-disk  is slightly above the mid-plane of M83 in the direction of the hidden nucleus, 
effectively shielding most of the light from it even
though the CO integrated intensities are not particularly high toward the dynamical center.

Alternatively, the visible nucleus may be a part of the original nucleus of M83 that
has moved away from the dynamical center of the galaxy. 
This would explain the lack of obvious disturbance in the M83 disk outside the nuclear ring.
Such a wandering nucleus may occur even without perturbation from outside the galaxy
\citep[and references therein]{Taga98}.

%%%%%%%%%%%%%%%%%%%%%%%%%%%%%%%%%%%%%%%%%%%%%%%

\section{Nuclear Starburst}
\begin{figure}
\epsscale{0.6}
\plotone{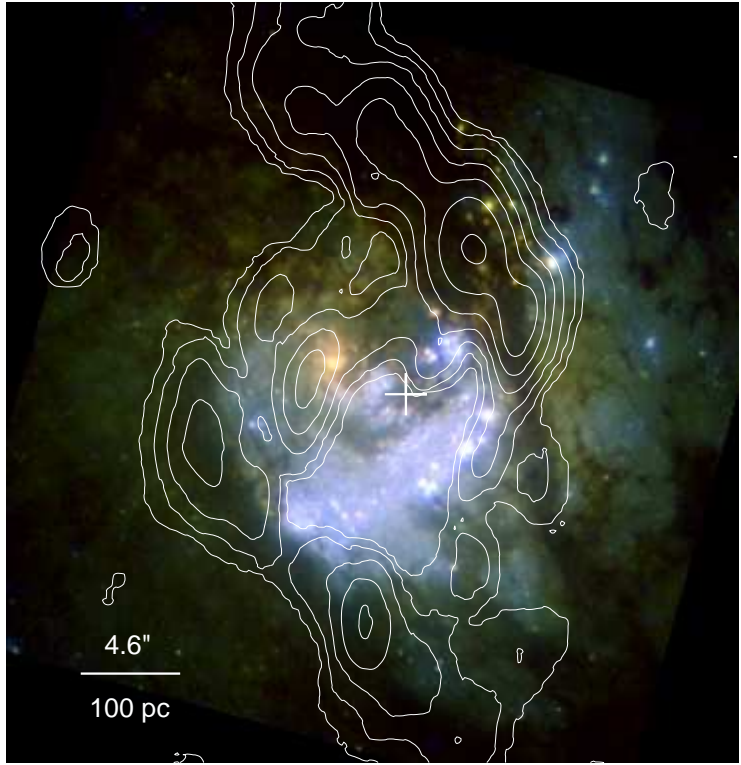}
\caption{Center of M83. 
SMA CO(3--2) contours, same as in Fig.\ref{fig.mom0}, are overlaid
on a three color composite made from archival HST/WFPC2 data.
F300W, F547M, and F814W images are used for blue, green, and red, respectively.
The visible nucleus is the orange peak, while the $K$-band centroid ($\approx$ dynamical center)
is marked with the cross.
The starburst arc is in white-blue and has little CO(3--2) emission on it.
\label{fig.hst}}
\end{figure}

The starburst in M83 is lopsided, mostly on the receding side of the dynamical center.
This is true not only  for the starburst arc (Figure \ref{fig.hst})
but also for other sites of active star formation more obscured by dust.
The latter  are $\sim$10\arcsec\ northwest of the visible nucleus,  
near the strongest peak of CO emission, 
according to radio and mid-infrared observations 
\citep{Turner94, Telesco93, Rouan96}. 

\placefigure{fig.hst}

The lopsided starburst may well be due to the dynamical effect of 
the off-center nucleus.
Lopsidedness ($m=1$ mode) is atypical in starburst morphology near the center of 
a barred galaxy, where bisymmetry (i.e., $m=2$ mode such as twin peaks, two-armed
spiral, and an oval ring) is most often observed.
Our observations established that the visible nucleus, be it an intruder or a wanderer, 
is off the dynamical center of the galaxy.
The new mass estimate of this component is $\sim$20 \% of the dynamical
mass in the central 300 pc of M83.
Thus the off-center nucleus must have had a 
significant, both positive and negative, influence on star formation in the area 
through perturbations to starforming gas. 
It is noteworthy that both the age of the starburst
and the dynamical timescale for the off-center nucleus 
are in the order of several Myr.
To summarize, the nuclear starburst of M83 owes much to the
bar-driven gas dynamics for accumulating molecular gas to the central 300 pc,
and appears to have taken the current shape under the strong influence of the double nucleus.

%%%%%%%%%%%%%%%%%%%%%%%%%%%%%%%%%%%%%%%%%%%%%%%

\acknowledgments

We are grateful to all people who worked hard to realize the SMA. 
We also thank P. T. P. Ho and J. M. Moran for helpful comments on the manuscript,
N. Z. Scoville for providing us with the MIR package before its release,  
and STScI for the archival HST images of M83.

%%%%%%%%%%%%%%%%%%%%%%%%%%%%%%%%%%%%%%%%%%%%%%%

%%%%%%%%%%%%%%%%%%%%%%%%%%%%%%%%%%%%%%%%%%%%%%%%%%
\clearpage

%%%%%

%%%%%

%%%%%

%%%%%%%%%%%%%%%%%%%%%%%%%%%%%%%%%%%%%%%%%%%%%%%%%%%


\clearpage

\begin{thebibliography}{}
\bibitem[Athanassoula(1992)]{Athanassoula92}
	Athanassoula, E. 
	1992, \mnras, 259, 345
\bibitem[Binney et al.(1991)]{Binney91}
	Binney, J., Gerhard, O. E., Stark, A. A., Bally, J., and Uchida, K. I. 
	1991, \mnras, 252, 210	
\bibitem[Cowan, Roberts, \& Branch(1994)]{Cowan94}
	Cowan, J. J., Roberts, D. A., and Branch, D.
	1994, \apj, 434, 128	
\bibitem[Crosthwaite et al.(2002)]{Crosthwaite02} 
	Crosthwaite, L. P., Turner, J. L., Buchholz, L., Ho, P. T. P., and Martin, R. N.  
	2002,  \aj, 123, 1892
\bibitem[Dumke et al.(2001)]{Dumke01}
	Dumke, M., Nieten, Ch., Thuma, G., Wielebinski, R., and Walsh, W.
	2001, \aap, 373, 853	
\bibitem[Elmegreen, Chromey, \& Warren(1998)]{Elmegreen98}
	Elmegreen, D. M., Chromey, F. R., and Warren, A. R.
	1998, \aj, 116, 2834	
\bibitem[Gallais et al.(1991)]{Gallais91}
	Gallais, P., Rouan, D., Lacombe, F., Tiph\`{e}ne, and Vauglin, I. 
	1991, \aap, 243, 309	
\bibitem[Genzel et al.(2001)]{Genzel01}
	Genzel, R., Tacconi, L. J., Rigopoulou, D., Lutz, D., and Tecza, M.
	2001, \apj, 563, 527
\bibitem[Handa et al.(1990)]{Handa90}
	Handa, T., Nakai, N., Sofue, Y., Hayashi, M., Fujimoto, M. 
	1990, \pasj, 42, 1	
\bibitem[Harris et al.(2001)]{Harris01}
	Harris, J., Calzetti, D., Gallagher III, J. S., Conselice, C. J., Smith, D. A.
	2001, \aj, 122, 3046
\bibitem[Ho, Moran, \& Lo(2004)]{Ho04}
	Ho, P. T. P., Moran, J. M., Lo, K. Y.
	2003, \apjl, {\it Letters}, in preparation
\bibitem[Israel \& Baas(2001)]{Israel01}
	Israel, F. P., and Baas, F. 
	2001, \aap, 371, 433	
\bibitem[Lindblad \& Lindblad(1994)]{Lindblad94}
	Lindblad, P. O., and Lindblad, P. A. B. 
	1994, in ASP Conf. Ser. 66, Physics of the gaseous and stellar disks of the galaxy, 
	ed. A. M. Fridman (San Francisco: ASP), 29
\bibitem[Malin \& Hadley(1997)]{Malin97}
	Malin, D. and Hadley, B.
	1997, Publ. Astron. Soc. Aust., 14, 52
\bibitem[Regan, Sheth, \& Fogel(1999)]{Regan99}
	Regan, M. W., Sheth, K., and Vogel, S. N. 1999, \apj, 526, 97
\bibitem[Rouan et al.(1996)]{Rouan96}
	Rouan, D. et al. 1996, \aap, 315, L141
\bibitem[e.g., Sakamoto, Baker, \& Scoville(2000)]{Sakamoto00}
	Sakamoto, K., Baker, A. J., and Scoville, N. Z. 
	2000, \apj, 533, 149
\bibitem[Sakamoto et al.(1999)]{Sakamoto99}
	Sakamoto, K., Scoville, N. Z., Yun, M. S., Crosas, M., Genzel, R., Tacconi, L. J.
	1999, \apj, 514, 68
\bibitem[Soria \& Wu(2002)]{Soria02}
	Soria, R., and Wu, K. 
	2002, \aap, 384, 99	
\bibitem[Taga \& Iye(1998)]{Taga98}
	Taga, M., and Iye, M. 1998, \mnras, 299, 111	
\bibitem[Telesco, Dressel, \& Wolstencroft(1993)]{Telesco93}
	Telesco, C. M., Dressel, L. L., and Wolstencroft, R. D. 
	1993, \apj, 414, 120
\bibitem[Thatte, Tecza, \& Genzel(2000)]{Thatte00}
	Thatte, N., Tecza, M., and Genzel, R. 
	2000, \aap, 364, L47 (TTG)
\bibitem[Thim et al.(2003)]{Thim03}
	Thim, F., Tammann, G. A., Saha, A., Dolphin, A., Sandage, A., Tolstoy, E., and Lbhardt, L.
	2003, \apj, 590, 256	
\bibitem[Tremaine(1995)]{Tremaine95}
	Tremaine, S. 1995, \aj, 110, 628	
\bibitem[Turner \& Ho(1994)]{Turner94}
	Turner, J. L. and Ho, P. T. P. 
	1994, \apj, 421, 122 	
\bibitem[Wada(1994)]{Wada94}
	Wada, K. 1994, \pasj, 46, 165
\bibitem[Walker, Mihos, \& Hernquist(1996)]{Walker96}
	Walker, I. R., Mihos, C. J., and Hernquist, L.
	1996, \apj, 460, 121

\end{thebibliography}
\end{document}